\documentclass[12pt,preprint]{aastex}

\begin{document}

\def\wisk#1{\ifmmode{#1}\else{$#1$}\fi}

\def\lt     {\wisk{<}}
\def\gt     {\wisk{>}}
\def\le     {\wisk{_<\atop^=}}
\def\ge     {\wisk{_>\atop^=}}
\def\lsim   {\wisk{_<\atop^{\sim}}}
\def\gsim   {\wisk{_>\atop^{\sim}}}
\def\kms    {\wisk{{\rm ~km~s^{-1}}}}
\def\Lsun   {\wisk{{\rm L_\odot}}}
\def\Zsun   {\wisk{{\rm Z_\odot}}}
\def\Msun   {\wisk{{\rm M_\odot}}}
\def\um     {$\mu$m}
\def\mic     {\mu{\rm m}}
\def\sig    {\wisk{\sigma}}
\def\etal   {{\sl et~al.\ }}
\def\eg     {{\it e.g.\ }}
 \def\ie     {{\it i.e.\ }}
\def\bsl    {\wisk{\backslash}}
\def\by     {\wisk{\times}}
\def\half {\wisk{\frac{1}{2}}}
\def\third {\wisk{\frac{1}{3}}}
\def\nwm2sr {\wisk{\rm nW/m^2/sr\ }}
\def\nw2m4sr {\wisk{\rm nW^2/m^4/sr\ }}

\title{Correcting the analysis of ``IR ANISOTROPIES IN SPITZER GOODS IMAGES..." by Cooray et al (2006)}

\author{
A. Kashlinsky\altaffilmark{1,2} } \altaffiltext{1}{Observational
Cosmology Laboratory, Code 665, Goddard Space Flight Center,
Greenbelt MD 20771 and SSAI} \altaffiltext{2}{e--mail:
kashlinsky@milkyway.gsfc.nasa.gov}

\begin{abstract}
We point out that in their analysis of the deep Spitzer images,
Cooray et al (2006) perform Fourier transform on maps which have
very few pixels left (only 20 to 30 percent). For such deeply cut
maps one cannot reliably compute large-scale map properties using
Fourier transforms. Instead the maps must be analyzed via the
correlation function, $C(\theta)$, which is immune to mask
effects. We find, when computing $C(\theta)$ for their maps, that
removing ACS/HST galaxies does not lead to appreciable change in
the correlation properties of the remaining diffuse emission. We
then demonstrate with simulations that the power spectrum of CIB
fluctuations {\it prior} to removal of the ACS galaxies reproduces
$C(\theta)$ in the maps from which the ACS galaxies have been
removed. This implies that these galaxies cannot be responsible
for the CIB fluctuations detected in Kashlinsky et al (2005,
2007), contrary to the claims of Cooray et al (2006).
\end{abstract}


Cosmic infrared background (CIB) anisotropies from early epochs
should contain also the contribution from the first stars and
galaxies (see Kashlinsky 2005 for review). In several attempts to
uncover this component, we have analyzed deep Spitzer exposures
identifying CIB fluctuations remaining after removal of galaxies
to fairly faint levels (Kashlinsky, Arendt, Mather \& Moseley
2005, 2007a,b). The initial findings were recently confirmed by
our analysis of the newly available Spitzer GOODS data
(Kashlinsky, Arendt, Mather \& Moseley 2007b), where we could
remove intervening galaxies to still fainter levels than in the
earlier study. The analyzed fields were clipped of sources so that
a reasonable fraction of pixels remained to allow a robust Fourier
analysis. This fraction must in practice be fairly high; e.g. even
when $\gsim 60\%$ of the original pixels are kept and the power
spectrum has a simple power-law behaviour, computations of the
power spectrum from Fourier transform may be misleading (Gorski
1994). Thus, for maps where fewer than $\sim 60 \%$ of the pixels
remained, Kashlinsky et al (2005) instead computed the diffuse
light correlation function, showing that its value decreases
little as the maps are clipped progressively deeper. Also note
that the CIB fluctuations signal was found to be present at all
four IRAC wavelengths with fairly high signal-to-noise
measurements at 3.6 and 4.5 micron, so that any interpretation of
the origin of that signal must explain all the wavelengths
simultaneously (Kashlinsky, Arendt, Mather \& Moseley 2007a) - not
just results derived from the 3.6 micron data.

Recently, Cooray et al (2006) have presented their own analysis of
the GOODS data at only 3.6 micron. In it they removed galaxies
identified at shorter wavelengths from the HST ACS observations
and claimed that the CIB fluctuations signal, whose power spectrum
was evaluated via Fourier transforming the maps, is significantly
diminished when this is done. {\it Note that in their Fourier
analysis they were left with only 30 \% (their map C) to 20 \%
(their map D) of the map pixels}. As is commonly known applying
Fourier analysis to such deeply cut maps would lead to spurious
results.

Indeed, Fourier analysis, which is meaningful only when the
masking effects leave the basis functions at least approximately
orthogonal, can lead to wrong results for the power spectrum,
$P(q)$, computed using it in such deeply cut maps. For such deeply
cut maps, one must use a complementary to $P(q)$ statistic, the
correlation function $C(\theta)=\langle \delta F(\vec{x}) \cdot
\delta F(\vec{x}+\vec{\theta})\rangle$ (e.g. Kashlinsky \&
Odenwald 2000, Matsumoto et al 2005) which is immune to masking
effects. This is why we presented this quantity in the
Supplementary Information to Kashlinsky et al (2005) for maps cut
at deeper levels leaving fewer than $\sim 65 \%$ of the map pixels
(see Fig. SI-4).

This quantity, $C(\theta)$, and not the Fourier transformed maps,
should also have been used by Cooray et al (2006) when analyzing
their maps in which fewer than $30\%$ of the pixels have remained.
Note that in this representation, the contributions of any white
noise (such as shot noise and/or instrument noise) component to
$C(\theta)$ drop off very rapidly outside the beam  (e.g. Smoot et
al 1992) and for the IRAC 3.6 \um\ channel contribute negligibly
to the correlation function at $\theta \gsim$ a few arcsec. Thus
$C(\theta)$ at these scales would reflect the clustering
component.

To see if the final maps of Cooray et al, when analyzed correctly,
indeed show less large scale CIB fluctuations, we have downloaded
their maps C and D from www.cooray.org where they are advertised
as publicly available. The CDF-S field does not have map C
available there and so we could not compare the large scale
correlations of the C and D maps. But we did that for the HDF-N
field. Fig. 1 shows the resultant $C(\theta)$ for their maps C
(when ACS galaxies are not removed) and D (when ACS galaxies are
removed). It shows that there is little difference between the two
maps in terms of the large-scale correlations and that, in fact,
map D has higher amplitude correlations (it also has a larger
variance than C).

For comparison, we also show the correlation function from Fig.
SI-4 of Kashlinsky et al (2005), which is essentially the same as
in the Cooray et al GOODS maps. This would suggest that the ACS
detected galaxies are not major contributors to the CIB
fluctuations contrary to the suggestions by Cooray et al (2006).

Indeed, one can verify with simulations that the correlation
function values in Fig.1 are consistent with the amount of large
scale power detected in KAMM1,2. In order to do this, we have
constructed ten realizations (computation of $C(\theta)$ is a very
CPU intensive procedure) of the CIB field with the power spectrum
corresponding to Map C (which includes the ACS galaxies) from Fig.
2 of Cooray et al., which as they note are consistent with the
measurements of KAMM1. We selected their central values; allowing
for the errors on power will make our conclusions stronger. Of the
ten computed $C(\theta)$, we show in Fig. 2 the realization with
the {\it second} smallest zero crossing and one with the {\it
second} largest, such that the frequency of any of these
realizations is $\sim 20\%$. The figure shows that the power
spectrum of diffuse light prior to removal of ACS galaxies
reproduces the correlation function after their removal within the
uncertainties determined by the field geometry and statistics.

I thank my collaborators Rick Arendt, John Mather and Harvey
Moseley for many useful discussions. This work is supported by NSF
AST-0406587 and NASA Spitzer NM0710076 grants.



\begin{figure}
\plotone{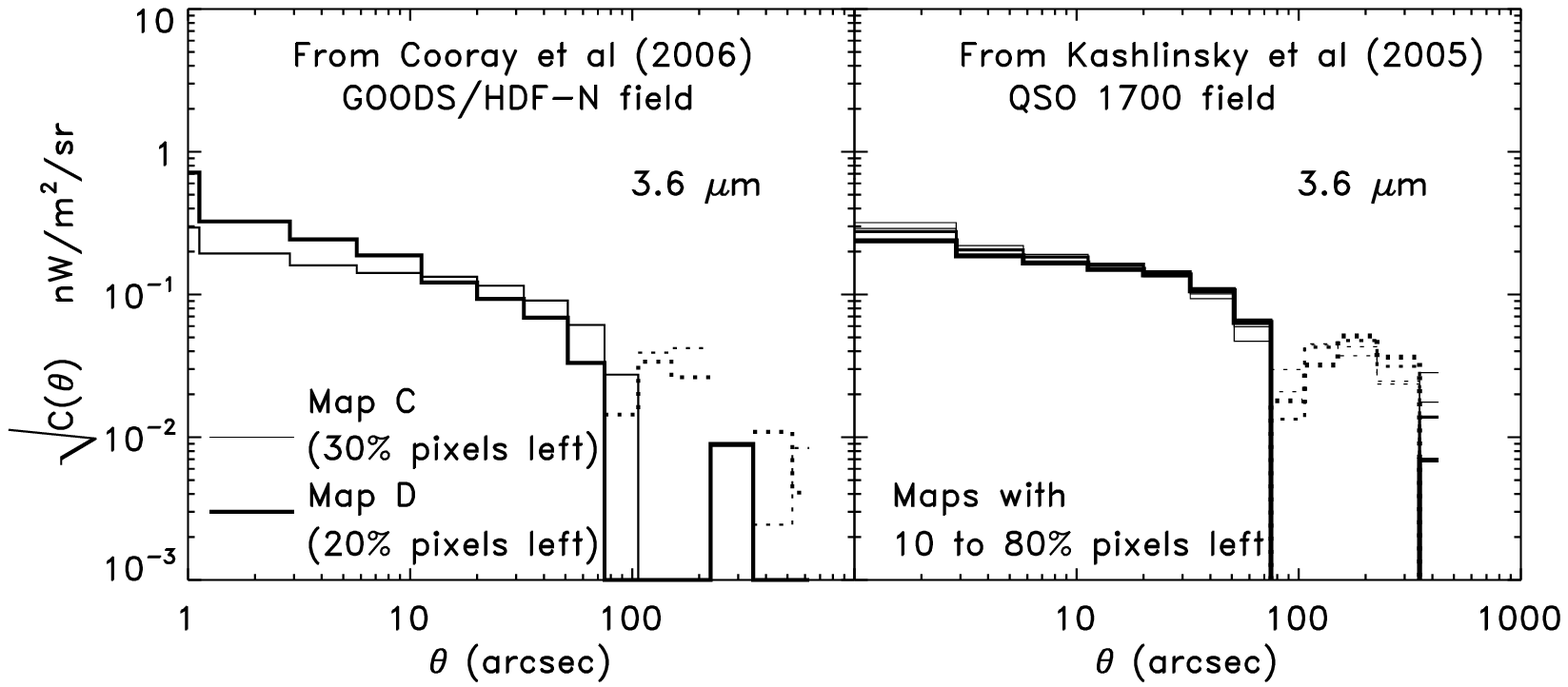} \caption{Solid lines correspond to
$\sqrt{C(\theta)}$ for $C>0$ and dotted lines to
$\sqrt{-C(\theta)}$ for $C<0$. Left: correlation function for the
deeply cut maps C (with ACS galaxies in) and D (with ACS galaxies
removed) from Cooray et al. (2006). Map C has only $\sim 30\%$
pixels left, while map D has even fewer pixels left of $\sim
20\%$. Right: For comparison, we show the correlation functions
from deeply cut maps for the QSO 1700 field from Kashlinsky et al
(2005). These maps are shown in the Supplementary Information
there and have between $\sim 10$ and $80\%$ pixels left, which
corresponds to the clipping parameters in the Kashlinsky et al
(2005) procedure of $N_{\rm cut}=2-4$ and $N_{\rm mask}=3-7$. As
one can see removing ACS galaxies by Cooray et al leaves
approximately the same correlation function as when keeping them
in.} \label{fig:shot-noise}
\end{figure}

\begin{figure}
\plotone{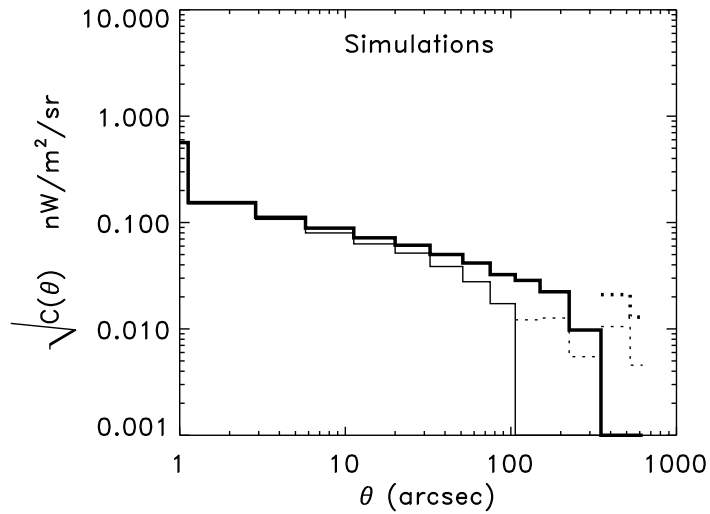} \caption{Results of simulations with 10
realizations of the CIB with the power spectrum given by the {\it
central} points of Map C from Fig. 2 of Cooray et al. The mask was
then applied and the correlation function evaluated. Of these ten
realizations we plot two: the realization with the {\it second}
smallest zero crossing and one with the {\it second} largest
zero-crossing. As in Fig. 1, the solid lines denote $C>0$ and
dotted correspond to the region of negative $C$. The figure shows
that it is very common to have large scatter in the values of
$C(\theta)$ at scales $>0.5-1$ arcmin for the geometries
corresponding to the clipped maps used in Cooray et al, but that
on smaller scales the correlation function reflects the power
correctly.} \label{fig:sims}
\end{figure}

\end{document}